# Entangled photon pairs produced by a quantum dot strongly coupled to a microcavity

R. Johne[1], N.A. Gippius[1,2], G. Pavlovic[1], D.D. Solnyshkov[1], I.A. Shelykh[1,3], G. Malpuech[1]

[1] *LASMEA, CNRS/University Blaise Pascal, 24 Avenue des Landais, 63177 Aubière, France*
[2] *A.M. Prokhorov General Physics Institute RAS, 119991 Moscow, Russia*
[3] *St. Petersburg State Polytechnical University, 195251, St-Petersburg, Russia.*

We show theoretically that entangled photon pairs can be produced on demand through the biexciton decay of a quantum dot strongly coupled to the modes of a photonic crystal. The strong coupling allows to tune the energy of the mixed exciton-photon (polariton) eigenmodes, and to overcome the natural splitting existing between the exciton states coupled with different linear polarizations of light. Polariton states are moreover well protected against dephasing due to their lifetime ten to hundred times shorter than that of a bare exciton. Our analysis shows that the scheme proposed can be achievable with the present technology.

Since the first treatment by Einstein, Podolsky, and Rosen (EPR) [1], quantum entanglement (EPR correlation) has been a widely discussed topic in theoretical physics [2]. The development of quantum information and communication [3] opened fascinating applications for EPR correlations, e.g. secret key exchange via the BB84 protocol [3,4]. A large number of theoretical proposals and experimental works on entangled photon sources based on these predictions have been published recently [5-11]. In this framework, semiconductor quantum dots (QDs) appear to be very good candidates to implement a solid state source of entangled photon pairs on demand, because of the rather long decoherence time of carriers in such structures. The basic idea which has attracted the strongest attention in the last years is to use the photon pairs produced by the decay of a biexciton [12]. The two possible decay paths for the biexciton in the ideal case can be distinguished only by measuring the polarisation of the two emitted photons of each cascade (fig.1-a). This system therefore appears to be a perfect solid state source of EPR pairs. However, even though QDs are often thought of as artificial atoms, they deviate from the ideal scheme because of the coupling between the exciton states with total angular momentum +1 and -1.

The anisotropic electron-hole interaction [13, 14] splits the exciton states into two modes linearly polarized along the crystallographic axis of the crystal, *H* and *V*. As a result, the photons emitted in the biexciton decay in a real QD are not polarization entangled, because they can be distinguished through their energy. Various solutions allowing to overcome this splitting have been proposed: carefully selected QDs [15, 16], filtering of the emission lines by the optical mode of a planar cavity [5], or application of an electric field [17,18,19]. Recently, two works have reported photon correlation measurements violating the Bell inequalities. The first one is based on the application of a magnetic field to control the position of the eigenstates [6]. The second one [8] is based on the idea that the two exciton emission peaks are overlapping and that the photons emitted in the overlap region are still polarisation entangled, which has been demonstrated thanks to a monochromatic detection scheme.

Another timely topic in modern physics is the strong coupling regime (SCR) in semiconductor microcavities. The SCR has been first achieved in planar cavities [20], where it allowed engineering of the dispersion and of the physical properties of the mixed eigenmodes of the cavity [21]. Recently, the SCR between an electronic excitation of a single QD and the optical mode of a photonic crystal has been reported by four different groups [22-26]. These implementations are opening a new research field where the precise control of the energy levels and the nature of the eigenmodes of the systems will be possible.

In this letter, we propose to use these new possibilities in order to achieve energy degenerate exciton-photon (polariton) modes, and thus to recover the ideal biexcitonic decay picture giving rise to polarisation entangled photon pairs.

A scheme depicting the idea of our proposal is sketched on the figure 1-b. The left part illustrates the biexciton decay in a realistic semiconductor QD. Starting from the biexciton state with the energy $E_{XX}$, the cascade passes through two intermediate exciton states indicated with $E_H$ for horizontal and $E_V$ for vertical polarisation, to reach the ground state (energy $E_G$.). The two bare exciton levels are split by an energy $\delta_X = E_H - E_V$ due to the exchange interaction. Because of this splitting the decay path can be distinguished by the photon energy.

In order to make the intermediate states degenerate, we propose to embed the QD within an anisotropic photonic crystal. Its cavity modes polarised $H$ and $V$ are split by an energy $\delta_C = E_C^H - E_C^V$, where ($E_C^{H,V}$) denotes the cavity mode energy for each polarization $H$ and $V$. Within the SCR, an exciton and a photon of a given polarisation couple and give rise to two similarly polarized polariton states. The energies of the four polariton states read as follows.

$$E_\pm^{H,V} = \frac{E_{H,V} + E_C^{H,V}}{2} \pm \frac{1}{2}\sqrt{(E_{H,V} - E_C^{H,V})^2 + 4\hbar^2\Omega_R^2} \,, \tag{1}$$

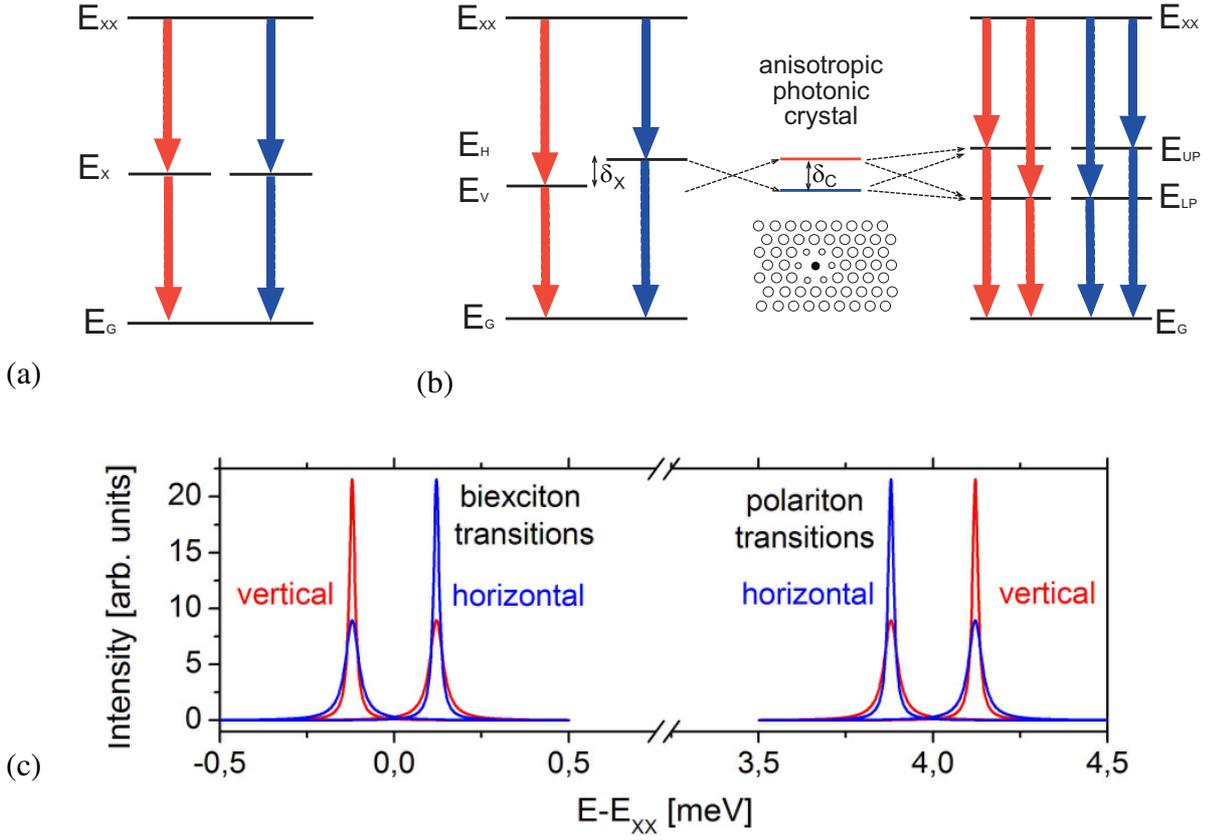

Fig. 1: (a) Schematic illustration of the ideal biexciton radiative cascade with $\delta_X = 0$. The different polarizations are indicated with red (vertical) and blue (horizontal). (b) Left figure shows the transitions in QDs with two split exciton levels $E_V$ and $E_H$. The middle part shows the eigenstates of the anisotropic photonic crystal and the scheme of the photonic crystal (open circles) as proposed in Ref.[28] with the embedded QD (black point). The sketch on the right shows the biexciton decay in the SCR. (c) shows the calculated photoluminescence spectra of the biexciton and polariton transitions. The different polarizations are indicated with red (vertical) and blue (horizontal).

$\Omega_R$ is the Rabi-splitting, which is taken equal for both polarisations [27]. The + sign stands for the upper polariton (UP) states and the - sign for the lower polariton (LP) states. The pairs of polariton states are degenerate ($E_\pm^H = E_\pm^V$) if $E_C^H = E_V$ and $E_C^V = E_H$, which means that each resonance for $V$ and $H$ polarized light is adjusted to the energy of the exciton state coupled to the perpendicular polarisation. In the same time, the biexciton transition is not strongly interacting with the cavity modes, because the binding energy of the biexciton is at least one order of magnitude larger than $\hbar\Omega_R$. This resonance can interact with another photonic mode, but we do not want to address this case here and we assume that the biexciton emission energy is not perturbed by the presence of the cavity. The right hand side of Fig.1-b shows the resulting distribution of the energy levels. There are four possible decay channels for the biexciton. The two decay paths using the UP as an intermediate state produce polarisation entangled photon pairs, which is also the case for the decay paths using the LP. This configuration is particularly original and probably useful, since it allows producing two independent EPR pairs. The corresponding photoluminescence spectra are shown in Fig.1-c.

The technological requirements for this scheme are however quite strong. The first condition is that $2\hbar\Omega_R > \delta_X$. The second condition is that the splitting between the optical modes is equal to the splitting between the QD modes with an opposite sign $\delta_X = -\delta_C$. The first condition is usually well fulfilled. In InAS based structures $\delta_X$ is of the order of 0.05-0.1 meV, whereas $2\Omega_R \approx 0.15$-0.25 meV. The second condition, because it is an equality, and because of the small value of $\delta_X$, seems quite demanding, and would, in practice, require the growth and study of many structures.

We therefore propose another configuration, conceptually less ideal, but which should allow an easier experimental implementation. We propose to use an anisotropic photonic crystal showing a splitting $\delta_C$ substantially larger than $\delta_X$. Neither the exact value, nor even the sign of $\delta_C$ play a crucial role in this scheme. This splitting should not be a problem, since it is difficult rather to fabricate photonic crystals without it. In Ref. [28], for instance, the splitting measured is about 0.5 meV for a cavity with quality factor Q>10000. Figure 2-a shows the eigenenergies versus the exciton-photon detuning $\delta_{C-X} = \frac{E_C^H + E_C^V}{2} - \frac{E^H + E^V}{2}$, keeping $\delta_X$ and $\delta_C$ constant. This kind of tuning of the exciton resonance energy can be performed experimentally, for example by changing the temperature of the sample [25,26]. We consider here the case where $\delta_X$ and $\delta_P$ have the same sign. For a wide range of detunings, the *H*-polarized LP and the *V*-polarized UP are almost degenerate. The decay channels of the biexciton are shown on the figure 2-b. The luminescence spectra in two polarizations for positive detuning are shown on the figure 2-c. The spectrum for each polarisation consists of two groups of two peaks. The group with the lower energy corresponds to the biexciton decay to the polariton states. The group with the higher energy corresponds to the decay of the polariton states toward the ground state. For each polarisation, the peak with the higher energy and the lower energy belong to the same decay cascade. The two central peaks belong to the same decay cascade as well. One can clearly see that the decay channel involving the H-polarized UP and the decay channel involving the V-polarized LP cannot be distinguished by energy measurements, but only by their polarisation.

As said before, this degeneracy can also be found if $\delta_X$ and $\delta_C$ have opposite signs. The Figure 3-a shows the eigenenergies versus $\delta_{C-X}$ in that case. The energy degeneracy now occurs at negative detuning between the LP states (*H* and *V)*, and at positive detuning between the UP states

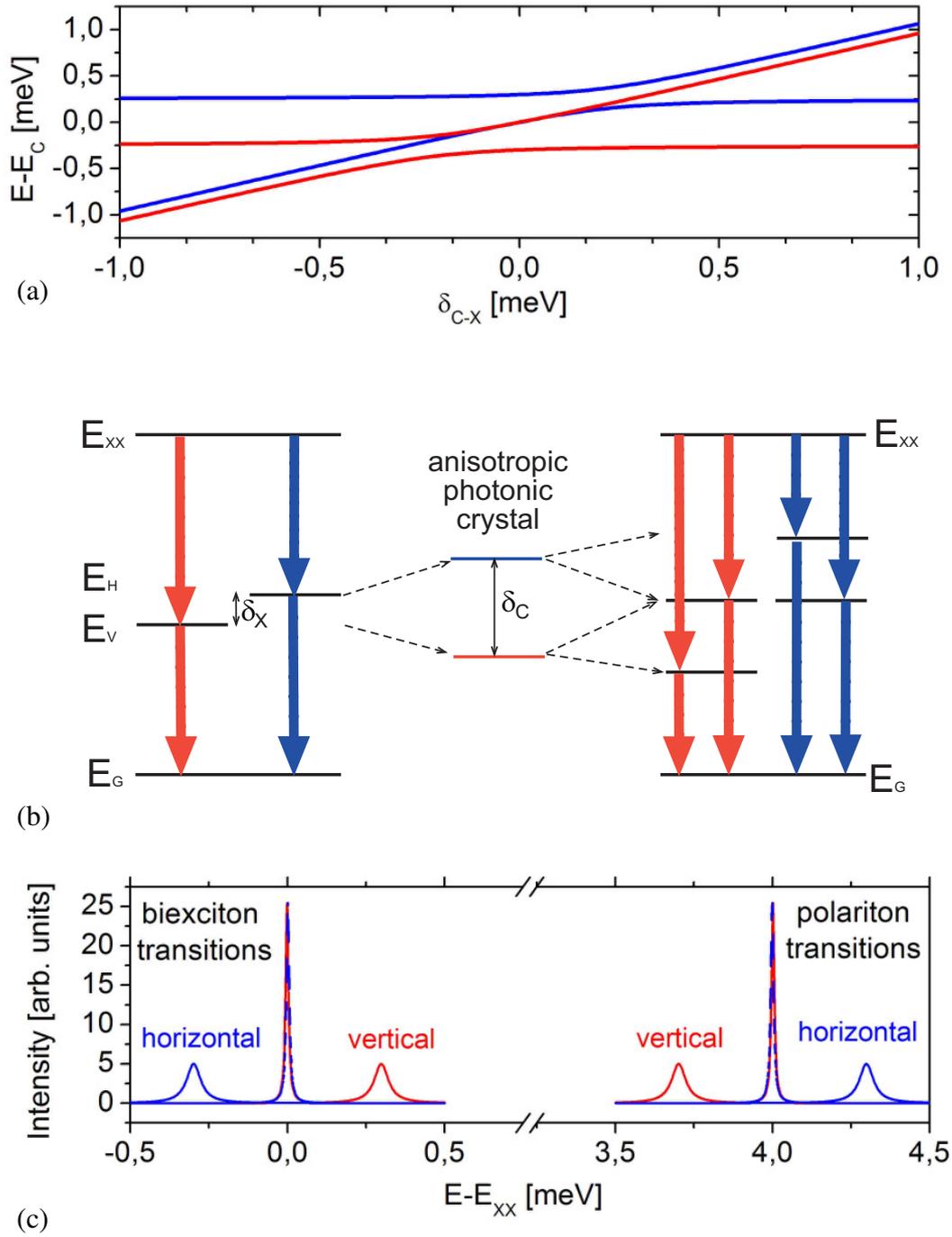

Fig.2: (a) Calculated energies of the polariton states for different detunings $\delta_{C-X}$ with $E_C = (E_C^H + E_C^V)/2$, $\delta_X$ =0.1 meV, $\delta_C$ = - 0.5 meV, and $2\hbar\Omega_R$ =0.22 meV. Different polarizations are indicated with red (vertical) and blue (horizontal). (b) Distribution of the energy levels for $\delta_{C-X}$ =0. (c) Photoluminescence spectra for both polarizations.

*(H and V)*. The decay channels of the biexciton for the negative detuning case are shown on the figure 3-b. The luminescence spectra for negative detunings are shown on the figure 3-c. Note the difference in the degeneracy of the peaks in figures 2-c and 3-c: in first case LP is degenerate with UP of different polarization, and in the second case LP is degenerate with LP.

The most important quality of the scheme we propose is the high degree of entanglement which can be expected for the photon pairs emitted. The general form of the wave-function of two photons generated by the biexciton decay can be written as [8]:

$$|\Psi\rangle = \left(\alpha_{LP}|p_H^{LP}\rangle + \alpha_{UP}|p_H^{UP}\rangle\right)|HH\rangle + \left(\beta_{LP}|p_V^{LP}\rangle + \beta_{UP}|p_V^{UP}\rangle\right)|VV\rangle, \quad (2)$$

where $\alpha_{LP}$, $\alpha_{UP}$, $\beta_{LP}$, $\beta_{UP}$ are the weights of the four possible transitions satisfying

$$|\alpha_{LP}|^2 + |\alpha_{UP}|^2 + |\beta_{LP}|^2 + |\beta_{UP}|^2 = 1. \quad (3)$$

The ket $|p_{H(V)}^{LP(UP)}\rangle$ is a coordinate part of the two photons wave packet for each polarization *H* and *V*, the kets $|HH\rangle$ and $|VV\rangle$ is a polarization part of a wavefunction corresponding to the two horizontally and vertically linear polarized photons respectively. The radiative decay of polariton states is governed by the cavity photon life time $\tau_C$ which is typically of the order of 10-20 ps — much faster than the exciton radiative decay. In the same time the transitions from one intermediate polariton state to another and the dephasing of these states have similar rates as in the bare exciton case, and the decay of the coherent intermediate state, which can play some important role in bare QDs systems [29], can be safely neglected here.

For the ideal case presented on figure 1-b, the density matrix of the system can be written as

$$\rho = \begin{pmatrix} |\alpha_{LP}|^2 + |\alpha_{UP}|^2 & 0 & 0 & \gamma \\ 0 & 0 & 0 & 0 \\ 0 & 0 & 0 & 0 \\ \gamma^* & 0 & 0 & |\beta_{LP}|^2 + |\beta_{UP}|^2 \end{pmatrix}, \quad (4)$$

where

$$\gamma = \alpha_{LP}\beta_{LP}^*\langle p_H^{LP}|p_V^{LP}\rangle + \alpha_{UP}\beta_{UP}^*\langle p_H^{UP}|p_V^{UP}\rangle. \quad (5)$$

where $\langle p_H^{UP}|p_V^{LP}\rangle$ and $\langle p_H^{LP}|p_V^{UP}\rangle$ have been taken equal to zero because they correspond to overlap between the wave functions of two photons of different energies. One can estimate the degree of entanglement using the Peres criterion [30], which, applied to our system, gives that the entanglement exists if and only if $|\gamma| > 0$. Two photons reach their maximally EPR correlated state for $|\gamma| = 1/2$ [28]. The polarization entangled photon pairs can be grouped by energy, one pair with

(a)

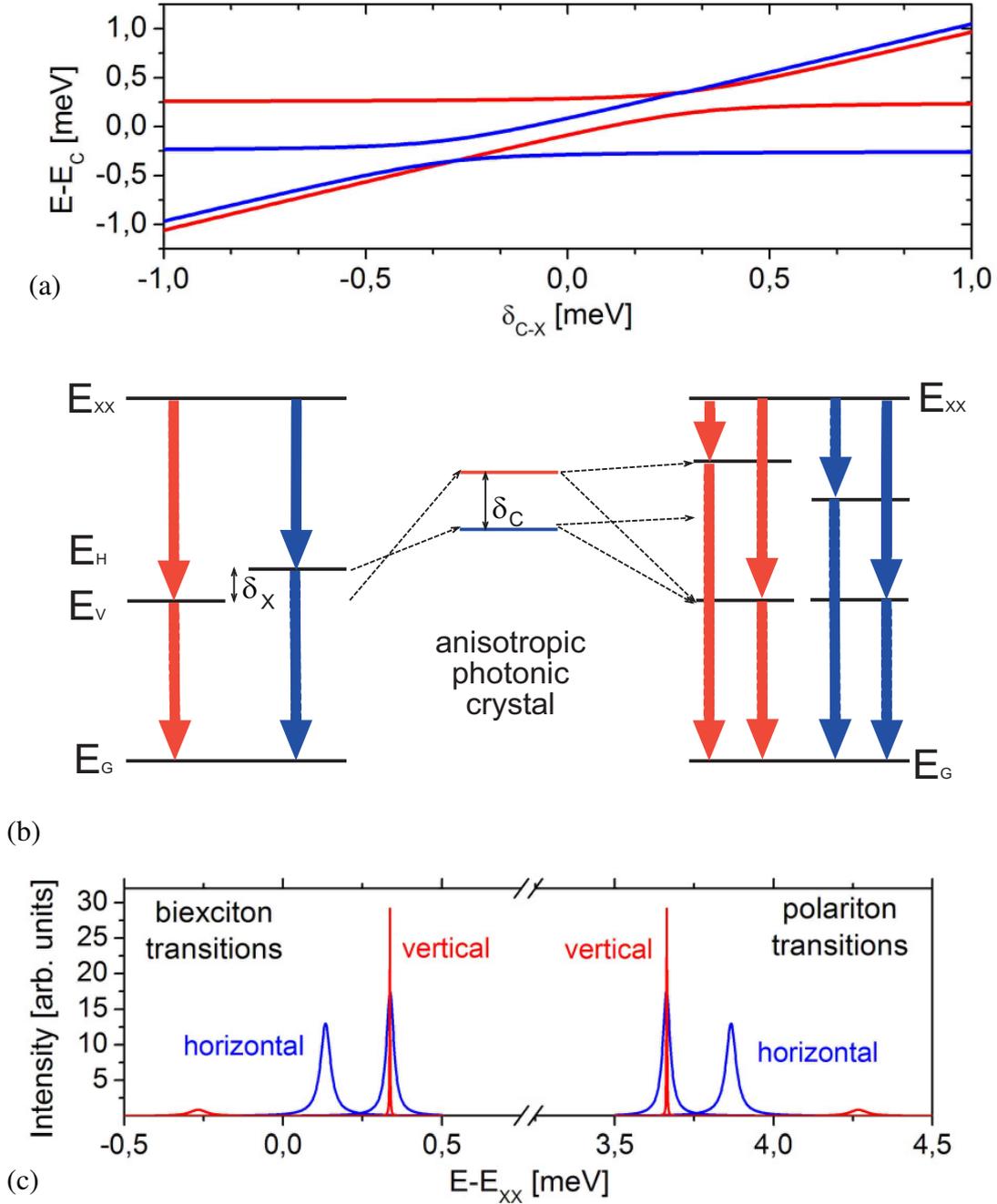

Fig.3:

(a) Calculated energies of the polariton states for different detunings $\delta_{C-X}$ with $E_C = (E_C^H + E_C^V)/2$, $\delta_X = $ -0.1 meV, $\delta_C = $ 0.5 meV, and $2\hbar\Omega_R = $0.22 meV. Different polarizations are indicated with red (vertical) and blue (horizontal). (b) Distribution of the energy levels for a positive $\delta_C$ and a negative $\delta_X$. (c) Photoluminescence spectra for both polarizations and a detuning $\delta_{C-X} = $ -0.275 meV.

$E_{LP}$ and one with $E_{UP}$. It is possible to define spectral windows of a detector to count only the entangled photons using one of the polariton levels, which is expressed as a projection of the wave packet. The wavefunction $|\Psi\rangle$ has to be recast as $P|\Psi\rangle/|P|\Psi\rangle|^2$. Fixing the spectral windows to the energy of the lower polariton branch $E_{LP}$ and $E_{XX}-E_{LP}$ leads to the disappearance of the upper branch terms. The diagonal terms of the density matrix (4) reduce to $|\alpha_{LP}|^2$ and $|\beta_{LP}|^2$. In this framework, the off-diagonal components of the density matrix thus yield

$$\gamma' = \frac{\alpha_{LP}\beta_{LP}^* \langle p_H^{LP} | P | p_V^{LP} \rangle}{|\alpha_{LP}|^2 \langle p_H^{LP} | P | p_H^{LP} \rangle + |\beta_{LP}|^2 \langle p_V^{LP} | P | p_V^{LP} \rangle}. \tag{6}$$

$\gamma'$ is equal to ½ if $\alpha_{LP} = \beta_{LP}$, which means that the two decay channels for the bi-exciton have the same amplitude, and the photon packets for different polarizations overlap perfectly.

To estimate the degree of entanglement of the presented schemes, it is necessary to determine the two photon function $|p_{H(V)}^{LP}\rangle$. Within the dipole and rotating wave approximation, the perturbation theory [8,31] gives

$$A_H^{LP} \equiv \alpha_{LP} \langle k_1, k_2 | p_H^{LP} \rangle = \frac{x_{ex}^{H,LP}\sqrt{\Gamma_{XX}} x_{ph}^{H,LP}\sqrt{\Gamma_{LP}^H}/2\pi}{(|k_1|+|k_2|-\varepsilon_{XX}^H)(|k_2|-\varepsilon_{LP}^H)}, \tag{7}$$

where $k_1$ and $k_2$ are the momenta of the photons ($\hbar, c = 1$), $\varepsilon_{XX}^H = E_{XX} - i\Gamma_{XX}$, and $\varepsilon_{LP}^H = E_{LP} - i\Gamma_{LP}^H$. A similar expression can be obtained for $A_V^{LP}$. The radiative width $\Gamma_{LP}^H = |x_{ph}^{H,LP}|^2/\tau_C$, where $x_{ph(ex)}^{H,LP}$ is the photon (exciton) Hopfield coefficient of the polariton state. On the other hand the radiative width of the biexciton-polariton transition is proportional to the exciton fraction of the polariton. Eq. (7) has a clear physical meaning: the wavefunction of a pair of emitted photons in the reciprocal space is a Lorentzian packet, whose width is given by the broadenings of the exciton and biexciton levels. Figure 4 shows the dependence of $|\gamma'|$ versus $\delta_{C-X}$, computed using

$$\gamma' = \frac{\iint dk_1 dk_2 A_H^{LP*} W A_V^{LP(UP)}}{\iint dk_1 dk_2 A_H^{LP*} W A_H^{LP} + \iint dk_1 dk_2 A_V^{LP(UP)*} W A_V^{LP(UP)}}, \tag{8}$$

where W represents the two spectral windows, and the upper branch transition amplitudes have been used to calculate $|\gamma|$ in scheme 2. The maximum values of $|\gamma'|$ for the scheme 3 is not optimal, due to the difference between the exciton and photon fractions of the degenerate polariton states. The asymmetry of the curves comes from the small lifetimes for negative detuning $\delta_{C-X}$. Consequently, the linewidth is larger than the energy difference between the two polariton states, which yields $\langle p_H^{LP} | P | p_V^{LP} \rangle > 0$. On the other hand, the degree of entanglement achieved within the schemes

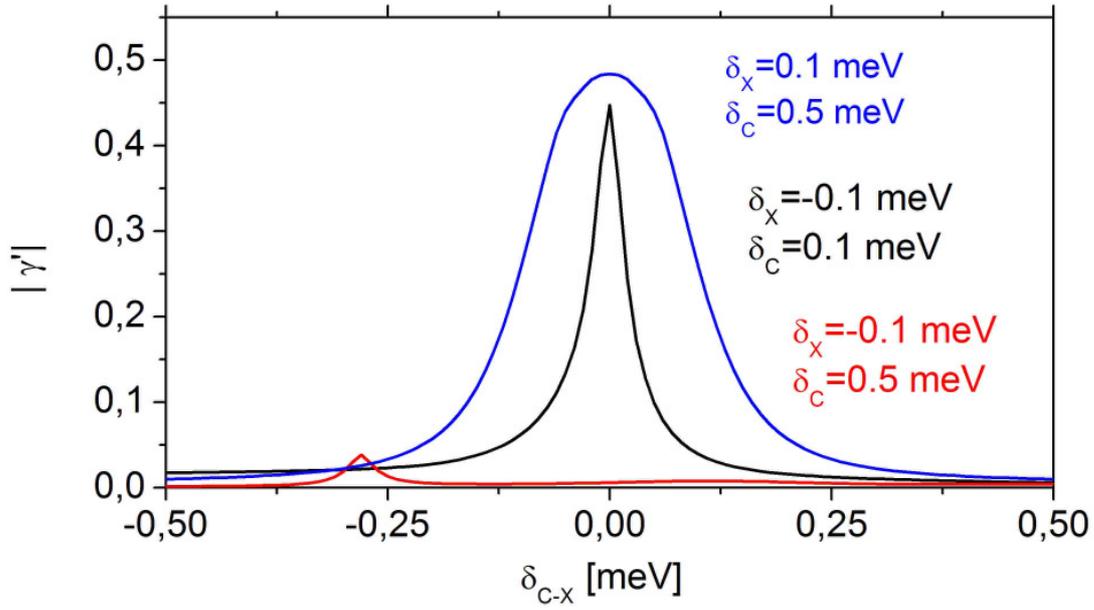

Fig.4:

Calculated $|\gamma'|$ versus $\delta_{C-X}$. Each color corresponds to a scheme from a different figure: 1 (black), 2 (blue), and 3 (red). Two detectors are placed at the energies $E_{XX}-E_{LP}$ and $E_{LP}$, with a spectral width of 0.2 meV.

proposed on figures 1 and 2 reaches almost the maximum value 1/2 at the intersection between the two polariton states, which makes these configurations quite favourable.

In conclusion, the SCR between light and matter allows engineering of the optical properties of quantum structures. We have shown that this approach allows transforming the split QD exciton levels into degenerate polariton levels. This scheme has moreover the advantage of rapidly decaying intermediate states, which are therefore extremely well protected from dephasing. We believe that our proposal can be realistically achieved experimentally using QDs embedded in photonic crystals, which is promising to implement a solid state source of EPR photon pairs for applications in quantum computing and quantum information processing.

The authors acknowledge financial support of the ANR Chair of Excellence Program and of the EU STIMSCAT FP6-517769 project.